\documentclass[aps,prb,twocolumn,showpacs,amsmath,amssymb]{revtex4}
\usepackage{dcolumn}
\usepackage{bm}
\usepackage{graphicx}
\usepackage{bm}
\begin{document}
\title{Effect of simultaneous application of field and pressure on magnetic transitions in La${_{0.5}}$Ca${_{0.5}}$MnO${_{3}}$}
\author{S. Dash, Kranti Kumar, A. Banerjee and P. Chaddah}
\affiliation{UGC-DAE Consortium for Scientific Research, University Campus, Khandwa Road, Indore-452001, Madhya Pradesh, India.}
\date{\today}
\begin{abstract}
We study combined effect of hydrostatic pressure and magnetic field on the magnetization of La${_{0.5}}$Ca${_{0.5}}$MnO${_{3}}$. We do not observe any significant effect of pressure on the paramagnetic to ferromagnetic transition. However, pressure asymmetrically affects the thermal hysteresis across the ferro-antiferromagnetic first-order transition, which has strong field dependence. Though the supercooling (T*) and superheating (T**) temperatures decrease and the value of magnetization at 5K (M$_{5K}$) increases with pressure, T* and M$_{5K}$ shows abrupt changes in tiny pressure of 0.68kbar. These anomalies enhance with field. In 7Tesla field, transition to antiferromagnetic phase disappears in 0.68kbar and M$_{5K}$ show significant increase. Thereafter, increase in pressure up to $\sim$10kbar has no noticeable effect on the magnetization.
\end{abstract}
\pacs{.}
\maketitle
Half doped perovskite manganites with generic formula of \textit{R${^{3+}_{0.5}}$A${^{2+}_{0.5}}$MnO${_{3}}$} have shown many interesting manifestation of electron correlation but continued to puzzle the community because of the drastic changes in physical properties arising form internal disorder  and/or external stimuli\cite{rao, dag1, tok1}. Coexistence of contrasting orders of almost similar energies and their tunablity with various parameters is considered to be responsible for the observed physical properties\cite{dag2, alok1}. The pressure-dependent behaviour of manganites close to half-doping, is being studied currently\cite{demko, sarkar}. However, these studies have been on samples that are close to, but below, half-doping and accordingly have ferromagnetic (FM) ground-state that is obtained directly from the high-temperature paramagnetic (PM) phase through a first order phase transition. In all these studies it has been observed that pressure favors FM phase, naively consistent with the expected increase in bandwidth, in that the transition temperature (T${_{C}}$) increases with increasing pressure. The hysteresis associated with the transition is also found to reduce with increasing pressure and the transition is reported to become second order above a critical pressure\cite{demko,sarkar}. 
 
On the other hand, prototype half doped manganites have charge ordered (CO)-antiferromagnetic (AFM) ground state and the intermediate bandwidth systems show AFM to FM transition followed by FM to PM transition on increasing T. In such half-doped systems, like Pr${_{0.5}}$Sr${_{0.5}}$MnO${_{3}}$ (PSMO), Nd${_{0.5}}$Sr${_{0.5}}$MnO${_{3}}$ (NSMO) and La${_{0.5}}$Ca${_{0.5}}$MnO${_{3}}$ (LCMO), the PM to FM transition is second order, and the FM to AFM transition is first order\cite{tok1, tok2, mori, cui, rue, yu, ashim}. The effect of pressure on the first order FM to AFM transition has been studied through resistivity measurements on NSMO and PSMO\cite{tok2, mori, cui, rue, yu}. For both these systems, it is shown that in the lower pressure range up to about 20 kbar the FM to AFM transition temperature (T${_{N}}$) increases with increasing pressure. This indicates that in this pressure range, coulomb interaction is enhanced at the cost of kinetic energy of the double exchange driven charge carriers with the increase in pressure for NSMO and PSMO.  

In this context, the intermediate bandwidth half doped manganite La${_{0.5}}$Ca${_{0.5}}$MnO${_{3}}$ may be considered as an attractive system to study the effect of pressure on the first order FM to AFM transition. Though this composition is very close to the phase boundary of CO-AFM and FM-M states, this extensively studied system is shown to have an AFM ground state that requires fields higher than 10T to convert it to an FM state\cite{alok2}. Recent magnetization measurements on polycrystalline LCMO shows PM to FM transition around 230K and FM to AFM transition around 120K on cooling in 1T\cite{lou,chaddah}. Studies on both samples\cite{lou,chaddah} show PM to FM transition to be second order while the FM to AFM transition shows large thermal hysteresis corresponding to a first order transition. Since magnetic field (H) is known to melt the CO state in this material, we study the effect of pressure (P) at various H. We explore how pressure affects the AFM state, especially in the presence of H. We find, in contrast to the reports on PSMO or NSMO, that pressure favors the FM state. We also find that, in the presence of H, there is an unexpected large effect at very small pressures.

The sample used here is the same as used in our earlier studies\cite{alok2, chaddah, alok3}. For completeness, we show in Fig 1 our earlier results M-H at 5K, and the zero-field cooled (ZFC), field-cooled cooling (FCC) and field-cooled warming (FCW) M-T curves in 1T field. The M-T behavior is similar to that seen by Loudon \textit{et al}\cite{lou}. Both Loudon \textit{et al.} and Banerjee \textit{et al.} concluded that at low-temperature the sample is an inhomogeneous mixture of FM and AFM regions. Moreover, Banerjee \textit{et al}\cite{alok3} showed that the FM regions persisting below the first order transition are a glass-like arrested state.

The magnetization measurement under external hydrostatic pressure have been carried out using a Cu-Be clamp type cell (easyLab Technologies) with a pressure volume of 1.9 mm diameter and 10 mm length attached in the SQUID magnetometer (M/s Quantum Design). The pressure was applied using a hydraulic press and was locked in at room temperature. Daphne mineral oil was used as the pressure-transmitting medium.  The maximum pressure that can be obtained in this cell is 10 kbar. The pressure value reported here was determined at low temperature by the known pressure dependence of the superconducting transition temperature of high-purity Sn, placed near the investigated sample. The error in determining the transition temperature is reflected as an error in the pressure values. It is found that this transition temperature of Sn remains reproducible as long as the pressure in the cell is not changed by changing the clamping. The measurement of the temperature dependence of magnetization in the range 5K$<T$ $<320K$ were performed in various magnetic fields by first applying H at 320K. The field was kept constant and M measured during cooling to 5K (FCC curve), and then again during warming to 320K (FCW curve). H was changed at 320K isothermally for measurements for the next value of H. After measurements at various values of H and T for a fixed pressure, the next value of pressure is applied to the pressure cell. Data is taken for seven values of pressure viz. 0 kbar, 0.68 kbar, 1.94 kbar, 2.93 kbar, 4.50 kbar, 6.99 kbar, and 9.12 kbar. It is to be noted that, after completion of measurements at 9.12 kbar, the pressure is released to 0 kbar, and in this condition the magnetization value is found to be the same as the initial zero pressure.

Recently, Kozlenko \textit{et al}\cite{koz} have shown that LCMO undergoes a structural transition from orthorhombic to monoclinic phase at a pressure near 150 kbar (at room temperature). Here we study LCMO only at lower P (below $\sim$10 kbar) and are thus looking only at effects on the electronic structure. Fig 2 shows FCC and FCW curves for M-T measured in seven different values of pressure (including zero) at different measurement field, H. Data was taken for various values of field, but we show data in figures 2 (a) to (d) for H ranging from 1 Tesla to 4 Tesla. It is rather noteworthy that there is hardly any P dependence, in this range of P, across the PM to FM transition, which is contrary to the recent observations on systems with FM ground states\cite{demko, sarkar}. However, we observed significant effect of pressure across the FM to AFM transition in the present system, which is shown to have AFM ground state. 
We identify the following three important parameters from the M-T curves for each value of P and H:

(i) T*, the temperature at which peak in M occurs during cooling run, i.e. the temperature at which the dM/dT changes sign in the respective cooling cycles,
 
(ii) T**, the temperature at which peak in M occurs during warming run i.e. the temperature at which the dM/dT changes sign in the respective warming cycles, and 

(iii) M$_{5K}$, the magnetization at 5K, which is a measure of the FM volume frozen (or trapped) in the CO-AFM matrix.
 
We plot these numbers as a function of P for each of these H in figure 3. Error bars represent the estimated uncertainties in P and in determining the temperatures at which dM/dT changes sign. We note from Fig. 3(a) that T** falls smoothly with increasing P for all H. This is in contrast to what is observed in NSMO and PSMO\cite{mori,cui,rue}, but is consistent with the expected increase in the bandwidth with increasing pressure. Fig. 3(b) shows that T* falls with rising P. This is again in contrast to what is observed in NSMO or PSMO\cite{mori,cui,rue}, but is consistent with the expected increase in bandwidth with increasing pressure. However, in this case, the fall is not smooth. Specifically, the fall is very sharp for the lowest P and the sharpness of the fall increases with measuring field, and is very sudden for 4 Tesla. This is an intriguing result and we shall discuss this later. However, this appears to have some relation with pressure dependence of M$_{5K}$ values. Fig. 3(c) shows that M$_{5K}$ rises with P but again the rise is not smooth. Specifically, the rise is very sharp for the lowest P, and the sharpness of this rise at the lowest non-zero pressure, increases as the measuring field is increased from 3T to 4T. Thus, both T* as well as M$_{5K}$ show similar anomalous behavior (sharp changes) at lowest non-zero pressure (0.68 kbar) and the anomaly becomes stronger as the measurement field is increased. 

The increasing effect of a minuscule pressure of 0.68 kbar with the measurement field is vividly demonstrated in Fig. 4. In Fig. 4(a) we shows the M-T curve while cooling and warming in field ranging between 1-7 Tesla measured without any pressure. This field range is much less than the field required to cause field induced AFM to FM transition at 5K ($\sim$ 10T) as evident in Fig. 1. However, the magnetization value at 5K increases significantly and FM-AFM transition becomes shallower with the increase in measuring field. This increase in the value of magnetization at 5K is much higher than the increase in M with H at 5K as shown in the M-H curve of Fig.1. Nonetheless, the first-order FM-AFM transition is evident even in 7T, in zero pressure. It is rather significant that a pressure of 0.68 kbar totally inhibited this FM-AFM transition at 7T, consequently there is considerable increase in magnetization at 5K, as depicted in Fig. 4(b). On the other hand, there is no distinguishable effect of further increase in pressure even by more than order of magnitude to 9.12 kbar. We are not aware of complete suppression of a transition by such a tiny hydrostatic pressure for any manganite or even any perovskite oxide. It appears that the magnetic field brings the system close to an instability and a small pressure causes drastic effect. 

We speculate that this could be due to the FM-AFM transition becoming a two-step process\cite{wak}. Recently, it is shown for the thin films of NSMO that the orbital ordering (OO) in the AFM phase changes with both temperature and field. In low fields, the temperature induced first-order FM to AFM transition occurs through an A-OO to CE-OO AFM phase as the temperature is decreased. Whereas in high field the transition is from FM to a phase separated state consisting of FM and CE-OO phases. It is shown that LCMO accrues glass like kinetically arrested FM phase when cooled in magnetic field across the FM-AFM transition\cite{lou,chaddah}. The fraction of this glass like FM phase increases with the increase in cooling field. It may be possible that this has some relation with the anomalous behavior observed for T* and M$_{5K}$ (Fig. 3). Since glass-like arrest is not known to interfere with the superheating spinodal, T** shows smooth decrease with pressure for all fields.

In conclusion, we report the simultaneous effect of magnetic field and hydrostatic pressure on temperature dependence of magnetization of half doped LCMO. No significant effect on the second-order PM to FM transition could be observed in the pressure range of $<$10 kbar, however, the lower temperature first-order FM-AFM transition is seriously affected, especially at the higher fields. The temperatures related to supercooling (T*) and superheating (T**) decrease and the magnetization value at 5K increases with the increase in pressure for all fields. However, decrease in T* and increase in M$_{5K}$ show sharp changes at the lowest non-zero pressure, sharpness of which increase with the field. Further, for 7T field the FM to AFM transition disappears, as a result M$_{5K}$ shows a significant increase in the tiny pressure of 0.68 kbar but thereafter even more than a order of increase in pressure value has no discernible effect on the temperature dependence of magnetization. We envisage that the sensitivity of spin/orbital order to the combined effect of field and pressure along with the ubiquitous presence of the glass-like arrested long-range ordered magnetic states in the half doped manganite may be responsible for such anomaly. However, systematic investigation of the simultaneous effect of pressure and field in the low-pressure regime is urgently needed to relate the microscopics with the important effects like phase-coexistence, bi-critical phase separation etc.

\begin{figure*}
	\centering
		\includegraphics[width=9cm]{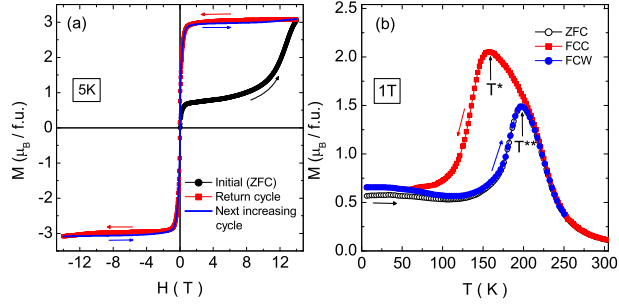}
		\caption{(Color online) Magnetization as a function H and T of La${_{0.5}}$Ca${_{0.5}}$MnO${_{3}}$. (a) M vs H at 5K after cooling from 320K in zero field. Initial field increasing cycle shows a field induced AFM to FM transition above 10T. However, the subsequent field decreasing or increasing cycles did not have any signature of reverse or forward transitions but the system appears as a soft-FM phase. (b) M vs T in 1T for La${_{0.5}}$Ca${_{0.5}}$MnO${_{3}}$ showing the second-order PM to FM transition followed by first-order FM to AFM transition with the decrease in T.}
\label{fig:Fig1}
\end{figure*}

\begin{figure*}
	\centering
		\includegraphics[width=9cm]{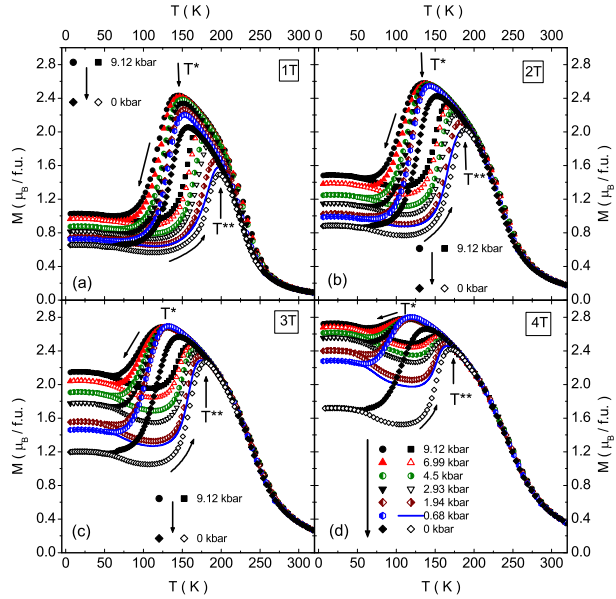}
		\caption{(Color online) Thermal hysteresis of magnetization of La${_{0.5}}$Ca${_{0.5}}$MnO${_{3}}$ under seven different pressures measured while cooling and warming in different fixed values of magnetic field. (a), (b), (c) and (d) shows magnetization in 1, 2, 3 and 4T respectively.}
\label{fig:Fig2}
\end{figure*}

\begin{figure*}
	\centering
		\includegraphics[width=7cm]{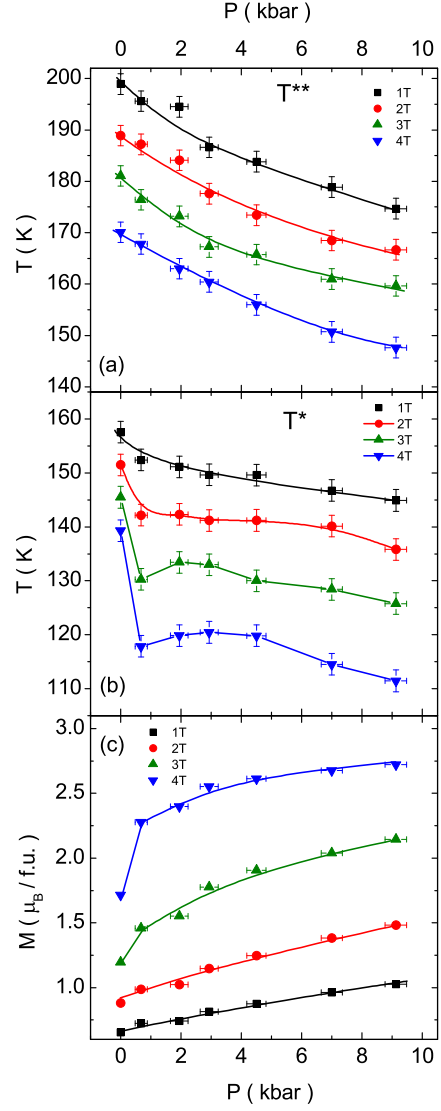}
		\caption{(Color online) Temperatures corresponding to the peak in M during superheating (T**), supercooling (T*) and values of M$_{5K}$ as a function of P for different measuring fields (H) as determined from the data of Fig. 2. Error bars corresponding to P and T are determined as described in the text. (a) T** vs. P for different H, (b) T* vs. P for different H and (c) M$_{5K}$ vs. P for different H.}
\label{fig:Fig3}
\end{figure*}

\begin{figure*}
	\centering
		\includegraphics[width=9cm]{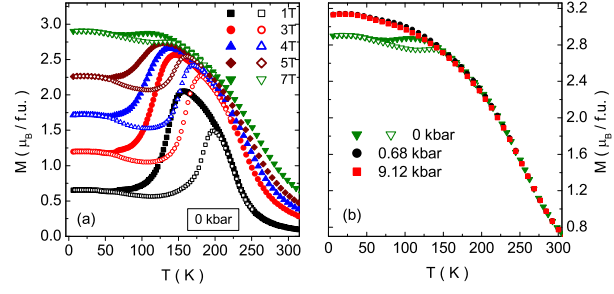}
		\caption{(Color online) (a) Magnetization of La${_{0.5}}$Ca${_{0.5}}$MnO${_{3}}$ while cooling (closed symbol) and warming (open symbol) in different H for zero external pressure. (b) The thermal hysteresis in M related to the FM to AFM transition observed in 7T disappears when a tiny pressure of 0.68 kbar is applied for the measurement in same field value. Thereafter, application of even 9.12 kbar pressure does not cause any noticeable change in M for the complete T range.}
\label{fig:Fig4}
\end{figure*}

\end{document}